\documentclass [11pt, prd,onecolumn,nofootinbib,notitlepage]{revtex4}
\usepackage{amsfonts,amsmath,amscd}
\usepackage{graphicx}
\usepackage{bbm}

\begin{document}

\title{%
Generally covariant formulation of Relative Locality in curved spacetime}

\author{F.\ Cianfrani}
\email{francesco.cianfrani@ift.uni.wroc.pl}\affiliation{Institute
for Theoretical Physics, University of Wroc\l{}aw, Pl.\ Maksa Borna
9, Pl--50-204 Wroc\l{}aw, Poland}
\author{J.\ Kowalski-Glikman}
\email{jkowalskiglikman@ift.uni.wroc.pl}\affiliation{Institute for
Theoretical Physics, University of Wroc\l{}aw, Pl.\ Maksa Borna 9,
Pl--50-204 Wroc\l{}aw, Poland}
\author{G.\ Rosati}
\email{giacomo.rosati@ift.uni.wroc.pl}\affiliation{Institute for
Theoretical Physics, University of Wroc\l{}aw, Pl.\ Maksa Borna 9,
Pl--50-204 Wroc\l{}aw, Poland}

\date{\today}
\small
\begin{abstract}
We construct a theory of particles moving in curved both momentum
space and spacetime, being a generalization of Relative Locality. We
find that in order to construct such theory, with desired
symmetries, including the general coordinate invariance, we have to
use non local position variables. It turns out that free particles
move on geodesics and momentum dependent translations of Relative
Locality are replaced with momentum dependent geodesic deviations.
\end{abstract}
\maketitle

\section{Introduction}

In the recent years the largely forgotten idea that momentum space
may have a nontrivial geometric structure, known under the name of
Born reciprocity \cite{Born:1949}, has been revived in many
different guises in the context of quantum gravity. It was noticed
in \cite{Majid:1999tc} that there is a one-to-one correspondence
between space-time noncommutativity, expected to be one of the
features of quantum gravity, and nontrivial geometric structures in
momentum space. This general observation is supported by explicit
calculations done in the context of gravity in 2+1 dimensions
\cite{Matschull:1997du}, \cite{Freidel:2005me}. Few years later it
was realized that many nontrivial features of Doubly Special
Relativity class of theories \cite{Amelino-Camelia:2000ge},
\cite{Amelino-Camelia:2000mn}, \cite{KowalskiGlikman:2001gp} can be
conveniently described in terms of the geometry of de Sitter
momentum space \cite{KowalskiGlikman:2003we}.

Relative Locality \cite{bob}, \cite{AmelinoCamelia:2011bm},
\cite{AmelinoCamelia:2011pe} is a theoretical framework that has its
roots in Born reciprocity. In this framework the momentum space is
brought to foreground. It is first observed that most, if not all,
physical measurements correspond, in fact, to momentum space data.
Second it is noticed that the emergence of a nontrivial geometry in
momentum space requires, as a prerequisite, the presence of a mass
scale. Such scale must be provided by a fundamental theory, and it
was assumed that there exists a regime of quantum gravity, in which
the length scale, the Planck length, is negligibly small, while the
mass scale, the Planck mass, remains finite.

In the couple of years that passed since Relative Locality was first
proposed the bulk of research investigated systems defined on
flat Minkowski spacetime. The question arises however if  curved
momentum space could coexists with a nontrivial geometry of
spacetime. 
This possibility is particularly intriguing from the phenomenological perspectives of DSR and Relative Locality, since many of the opportunities that have been proposed in the recent years rely on tests of (Planck-scale) deformation of kinematics of particles coming from cosmological distances~\cite{dsrphen},\cite{unoEdue}. The incredibly small size of such effects ($1/M_p \sim 10^{-19} \text{GeV}$) could be within the reach of present observations thanks to the huge amplification provided by the cosmological distances. In such context it is clear that the effects of spacetime curvature cannot be ignored.
Some preliminary results on the interplay between spacetime expansion and relativity of locality have been presented in the paper~\cite{AmelinoCamelia:2012it}, for the case of a de Sitter-like spacetime expansion.

Recently two of the present authors proposed the action of a
particle moving in curved spacetime, whose geometry is given by the
tetrad $e^a_\mu(x)$, with curved momentum space, provided by the
tetrad $E_a^\alpha(p)$~\cite{Kowalski-Glikman:2013xia}. While the
framework presented in~\cite{Kowalski-Glikman:2013xia} reproduces the correct action in both
the complementary limits of flat spacetime/curved momentum space and
curved spacetime/flat momentum space, the theory described by the
action proposed in~\cite{Kowalski-Glikman:2013xia} is not manifestly invariant under general
coordinate transformations (at least in their classical form). This,
by itself, may not be very problematic, because the scale governing
the effects related to this loss of invariance would be, in this
theory, presumably of order of the Planck scale,  and therefore they may not lead to relevant
phenomenological consequences. However, from the theoretical
perspective, one may find annoying the fact that the violation of
the general coordinate invariance would lead to the presence of a
preferred spacetime coordinate system, in which the particle action
has the form proposed in~\cite{Kowalski-Glikman:2013xia}. The
problem would be then to find out what this coordinate system is: is
it indeed the system of Cartesian coordinates in Minkowski space, in
the case of flat spacetime, as implicitly assumed in the
construction of Relative Locality? And, more importantly, how to
find such system, and the form of the particle action, in the case
of an arbitrary spacetime?

One possibility to avoid this conflict could be to look for a
generalized  class of coordinate transformations under which the
action presented in~\cite{Kowalski-Glikman:2013xia} is still
invariant. This would lead inevitably to coordinate transformations
mixing spacetime and momentum space. Taking into account that the
existence of such kind of transformations is not guaranteed, it may
also be conceptually compelling to explore the implications of
having a theory of both spacetime and momentum space curved, in
which invariance under general coordinate transformation is lost at
the Planck scale. While we postpone these analyses to future
studies, we feel however that the option of a coordinate invariant
theory is still the most desirable road to pursue in the search of
generalizing relative locality to curved spacetime.

Thus, in this paper we present a novel formulation of the action for
particles in both curved spacetime and momentum space, which is
manifestly invariant under general coordinate invariance. We also
show that in this theory the particles are described to move along
worldlines coinciding with  the standard spacetime geodesics. In the
next section we show how one can construct such an action for free
particles. In section 3 we discuss symmetries of so defined theory. 
In section 4 we show how, starting from this action, one
can also introduce particle interactions in the spirit of Relative
Locality. 
The final section is devoted to discussion.

When this work was being completed we learned about an interesting
complementary results presented in \cite{Freidel:2013xwa}.

\section{Construction of the action}

The action of a free relativistic particle with curved momentum
space has the form reciprocal, in a sense, to the one of the
standard free relativistic particle moving in curved spacetime, with
flat momentum space. The Lagrangian of the latter reads
\begin{equation}\label{1}
    L =  \dot{x}^\mu(\tau)\, e^a_\mu(x(\tau))\, p_a(\tau) - N(\eta^{ab} p_a p_b
    - m^2)\,,
\end{equation}
where $x^\mu(\tau)$ is the position of the particle at time $\tau$,
$p_a(\tau)$ is the particle momentum, and $e^a_\mu$ is the tetrad,
characterizing the geometry of spacetime
\begin{equation}\label{1a}
e^a_\mu e^b_\nu\eta_{ab} = g_{\mu\nu}\,,\quad e^a_\mu e^b_\nu
g^{\mu\nu} =\eta^{ab}\,.
\end{equation}
 Finally $N$ is the Lagrange multiplier
enforcing the mass-shell constraint $p^2=m^2$. To write the
Lagrangian~(\ref{1}) we use two kind of indices: the curved
spacetime index $\mu$ and the index $a$ related to the orthonormal
coordinate system in the ambient Minkowski space, to which the
tetrad $e^a_\mu$ maps. It can be checked by direct calculation that
the Euler-Lagrange equations following from (\ref{1}) reduce, after
solving for $p$, to the standard geodesic equation.

The action~(\ref{1}) is the first order form of the better known second
order action. The latter can be obtain from~(\ref{1}) by solving the
momentum equation of motion and substituting it back to~(\ref{1});
as a result one obtains
$$
L\sim g_{\mu\nu}(x(\tau))\,\dot{x}^\mu(\tau)\,\dot{x}^\nu(\tau)\,.
$$

Before we proceed, let us recall the basic properties of the
tetrads. It follows from the defining equation that tetrads are
spacetime vectors, transforming under diffeomorphisms as
\begin{equation}\label{1b}
    \delta_\xi e^a_\mu(x) = \xi^\nu\partial_\nu e^a_\mu +
    e^a_\nu\partial_\mu\xi^\nu\,.
\end{equation}
Moreover the relations (\ref{1a}) are invariant under infinitesimal
local Lorentz transformations
\begin{equation}\label{1c}
    \delta_\lambda e^a_\mu(x) =\lambda^a{}_b\, e^b_\mu\,,\quad
   \lambda^{ab}=-\lambda^{ba}\,.
\end{equation}
These two symmetries commute $[\delta_\xi, \delta_\lambda]\,
e^a_\mu=0$. In addition, assuming vanishing torsion, the tetrad
satisfies
\begin{equation}\label{1d}
    \partial_{[\mu} e^a_{\nu]} + \omega_{[\mu}{}^a{}_{b}\,
    e^b_{\nu]}=0\,,
\end{equation}
where $\omega$ is a gauge field for local Lorentz symmetry,
transforming as
\begin{equation}\label{1e}
    \omega_\mu' =\Lambda^{-1}\partial_\mu\Lambda + \Lambda^{-1}\omega_\mu\Lambda
\end{equation}
where $\Lambda$ is the matrix of a finite Lorentz transformation
$\Lambda=\exp(\lambda^{ab}T_{ab})$, with $T_{ab}$ being the matrix
generators of the Lorentz group. Finally the tetrads satisfy the 
tetrad postulate, according to which they are covariantly constant
\begin{equation}\label{1f}
\partial_{\mu} e^a_{\nu} + \omega_{\mu}{}^a{}_{b}\,
    e^b_{\nu}-\Gamma^\alpha_{\nu\mu}\, e^a_{\alpha} =0\,.
\end{equation}

It follows from the the Born reciprocity idea that the kinetic part of
the relativistic particle Lagrangian in the case of curved momentum
space should look like (\ref{1}) with the roles of $x$ and $p$
exchanged. One also has to replace the mass-shell condition with its
curved momentum space counterpart ${\cal C}(p) - m^2=0$, where
${\cal C}(p)$ is the square of the geodesic distance between the point
with coordinates $p^\alpha$ and the origin of the momentum space,
obtaining as a result \cite{AmelinoCamelia:2011bm},
\cite{Kowalski-Glikman:2013rxa}
\begin{equation}\label{2}
    L^{RL} = \dot p_\alpha(\tau)\, E^\alpha_a(p(\tau))\, x^a(\tau)
    + N({\cal C}(p) - m^2)\,.
\end{equation}
It should be recalled at this point that although the action
(\ref{2}) looks much more complex than the one of the standard
relativistic particle moving in flat space, the equations of motion
following from it are remarkably similar: they say that momentum
$p_\alpha$ and velocity $\dot x^a$ are both constant. What makes the
action (\ref{2}) different from its standard counterpart is the
relation between momentum and velocity, which becomes in the case of
(\ref{2}) highly nonlinear. Moreover, when interactions between
particles are introduced, the effects of relative locality starts being visible.

Generalizing these considerations, we want now to construct an
action for a particle with momentum space and spacetime both
possessing nontrivial geometries. Of course, we want the action
to reproduce the two limiting cases of flat momentum space/flat
spacetime discussed above. We require moreover the new action to be
still manifestly invariant under general coordinate transformations.
In order to meet these requirements, we introduce non local
variables which we denote $X^a$. We will
describe their construction in the following subsection.

\subsection{Non-local variables}

 Let us denote by
$\Gamma$ the $C^\infty$ curve $x=x(\tau)$ for $\tau\in[t_1,t_2]$ and
consider the sub-curves $\Gamma_\tau:\,x=x(\sigma)$ with
$\sigma\in[t_1,\tau]$ which coincide with $\Gamma$ up to $x(\tau)$.

As discussed above, given the background spacetime metric
$g_{\mu\nu}$, there exist a whole family of tetrads satisfying eq.\
(\ref{1a}), which differ from one another by action of local Lorentz
transformations. We can now gauge fix the local Lorentz
transformations in such a way that the Lorentz connection
$\omega^a_{\mu}{}_{b}$ vanishes along a given curve $x=x(\tau)$ and
then it follows from (\ref{1d}) that on this curve we can construct
a tetrad $\bar{e}^a_\mu$ with $\bar e^a_\mu \bar e^b_\nu\eta_{ab} =
g_{\mu\nu}$ such that
\begin{equation}\label{3}
(\partial_\nu \bar{e}^a_\mu(x)-\partial_\mu
\bar{e}^a_\nu(x))|_{x=x(\tau)}=0,
\end{equation}

The existence of the Lorentz connection with these properties can be proved as follows. 
We first show that the component of
$\omega_\mu$ along the worldline $\Gamma$, $\omega_\tau^a{}_{b}\equiv
\dot x^\mu \omega_\mu^a{}_{b}$ can be gauge fixed to zero. To this
end we have to solve the equation (cf.\ (\ref{1e}))
$0=\Lambda^{-1}\dot\Lambda +\Lambda^{-1}\omega_\tau\Lambda$. But
this equation is solved by a time ordered Wilson line (holonomy),
$\Lambda = T\exp\left(-\int d\tau \omega_\tau\right)$. Having fixed
$\omega_\tau=0$ we are left with the gauge transformations that are
constant along $\Gamma$. Let us now consider an arbitrary constant
time surface, corresponding to some particular value of the
parameter $\tau$ on $\Gamma$.\footnote{We assume that the worldline
is timelike, but an analogous construction works in the case of null
worldlines.} Then in the vicinity of the point in which the worldline
crosses the surface we have $\Lambda(\tau, x^i) = \Lambda^{(0)} +
\Lambda^{(1)}_i(\tau)\, x^i + O(x^2)$. This is sufficient freedom to
gauge fix to zero the spacial components of Lorentz connection
$\Lambda^{(0)}{}^{-1}\Lambda^{(1)}_i(\tau)=-\Lambda^{(0)}{}^{-1}\omega_i(\tau)\Lambda^{(0)}$.
This can be done for any $\tau$ and therefore all the components of
Lorentz connection can be gauge fixed to zero along a curve. After
gauge fixing $\omega_i=0$ on $\Gamma$, we are left with the gauge
freedom $\Lambda(\tau, x^i) = \Lambda^{(0)}  + O(x^2)$. Therefore,
once we gauge fix the connection along one curve, in general we
cannot do the same for another curve in its small neighborhood.

The tetrads $\bar{e}^a_\mu$ (\ref{3}) are determined modulo a global
Lorentz transformation $\Lambda^{(0)}$, which  can be used to fix
them equal to an arbitrary tetrad $e^a_\mu$ at one point of
$\Gamma$,
\begin{equation}\label{4}
\bar{e}^a_\mu(x(\bar{\tau}))=e^a_\mu(x(\bar{\tau}))
\end{equation}
Since the construction of the tetrads $\bar{e}^a_\mu$ reminds the
one of Fermi coordinates, in what follows we will call them Fermi
tetrads.

With these prerequisites we are ready to define the nonlocal variable
$X^a$ as
\begin{equation}\label{5}
    X^a(\Gamma; x(\tau)) = \int_{\,\,\Gamma_\tau}{}\, d\sigma\, \bar{e}^a_\mu(x(\sigma))\, \frac{dx^\mu}{d\sigma}=\int_0^\tau d\sigma\, \bar{e}^a_\mu(x(\sigma))\,
    \dot{x}^\mu\,,
\end{equation}
The variable $X^a(\Gamma; x(\tau))$ depends in general on the curve
$\Gamma$ along which it is calculated.

\begin{figure}[h!]
\includegraphics[scale=0.4]{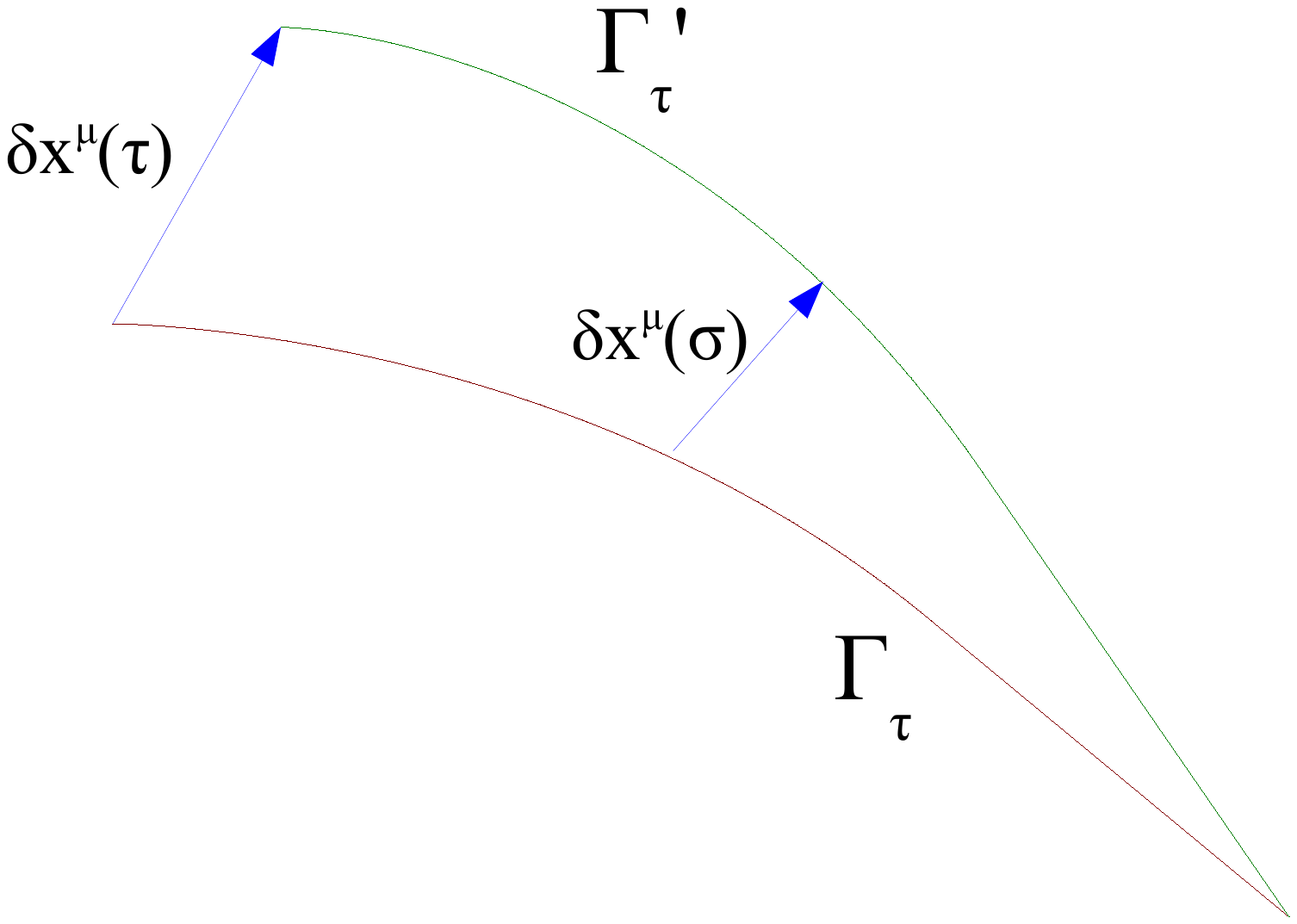}
\caption{The curves
$\Gamma_\tau:x^\mu(\sigma)$ and
$\Gamma'_\tau:x^{\mu}(\sigma)+\xi^\mu(\sigma)$.} \label{curves}
\end{figure}

\noindent

In order to calculate the variation of the action below, we will have
to evaluate the difference between variables $X^a$ calculated along
different curves $\Gamma$ and $\Gamma'$ lying infinitesimally close
to each other, with appropriate Fermi tetrads associated with each
of them. Thus
\begin{equation}
\delta{X}^a=\int_{\Gamma'}\bar{e}^a_{\mu\,\Gamma'}(x+\delta
x)(\dot{x}^\mu+\delta \dot{x}^\mu)\,d\sigma-
\int_{\Gamma}\bar{e}^a_{\mu\,\Gamma}(x)\dot{x}^\mu\,d\sigma\,,
\end{equation}
where $\bar{e}^a_{\mu\,\Gamma}$ denotes a tetrad field, defined in
spacetime, which becomes a Fermi tetrad on the curve $\Gamma$. The
variation of the tetrads $\bar{e}^a_\mu$ can be decomposed into two
parts:
\begin{equation}
\delta\,\bar{e}^a_{\mu\,\Gamma}(x)=\delta_1\,\bar{e}^a_\mu(x)+\delta_2\,\bar{e}^a_\mu(x),
\end{equation}
with
\begin{equation}
\delta_1\,\bar{e}^a_{\mu\,\Gamma}(x)=\bar{e}^a_{\delta\mu\,\Gamma'}(x+\delta
x)-\bar{e}^a_{\mu\,\Gamma}(x+\delta x),\label{delta1}
\end{equation}
and
\begin{equation}
\delta_2\,\bar{e}^a_{\mu\,\Gamma}(x)=\bar{e}^a_{\mu\,\Gamma}(x+\delta
x)-\bar{e}^a_{\mu\,\Gamma}(x)=\delta x^\nu
\bar{e}^a_{\mu,\nu\,\Gamma}.
\end{equation}
In order to evaluate the expression (\ref{delta1}) we need to
compare the Fermi tetrads associated with different curves. Since
they are both tetrads of the same spacetime metric, there exist a
local Lorentz transformation $\Lambda$ relating them
\begin{equation}
 \bar{e}^a_{\mu\,\Gamma'}(x)=\Lambda^a_{\,b}(x)\,\bar{e}^b_{\mu\,\Gamma}(x),
\end{equation}
and the associated Lorentz connections  are related by the local
Lorentz gauge transformation
\begin{equation}
\bar{\omega}_{\mu\,
\Gamma'}(x)=\Lambda^{-1}(x)\bar{\omega}_{\mu\,\Gamma}(x)\Lambda(x)+\Lambda^{-1}(x)\,\partial_\mu\Lambda(x).
\end{equation}
We know that the Lorentz connection $\bar{\omega}_{\Gamma'}$
vanishes on $\Gamma'$ and $\bar{\omega}_{\Gamma}$ vanishes on
$\Gamma$; thus we have
\begin{align}
0&=\bar{\omega}_{\mu\,\Gamma'}(x+\delta
x)=\Lambda^{-1}\bar{\omega}_{\mu\,\Gamma}\Lambda(x+\delta
x)+\Lambda^{-1}\,\partial_\mu\Lambda(x+\delta
x)=\nonumber\\
&=\Lambda^{-1}(x)\left(\delta x^\nu\bar{\omega}_{\mu,\nu\,\Gamma}(x)
\right)\Lambda(x)+\Lambda^{-1}\,\partial_\mu\Lambda(x+\delta x)\,.
\end{align}

For an infinitesimal local Lorentz gauge transformation
$\Lambda(x)\simeq I+\lambda(x)$, keeping the leading order terms we
obtain
\begin{equation}
0=\delta x^\nu\bar{\omega}_{\mu,\nu\,\Gamma}(x) +\partial_\mu
\lambda(x)\,,
\end{equation}
from which, multiplying by $\dot x^\mu$ and adding a term
proportional to $\bar{\omega}^{ab}_{\mu\,\Gamma}=0$, we get
\begin{equation}\label{19}
\frac{d\lambda^{ab}}{d\sigma}=-\delta
x^\nu\,\bar{\omega}^{ab}_{\mu,\nu\,\Gamma}\,\dot{x}^\mu-
\bar{\omega}^{ab}_{\mu\,\Gamma}\,\delta \dot{x}^\mu=
-\frac{d}{d\sigma}(\delta
x^\mu\,\bar{\omega}^{ab}_{\mu\,\Gamma})-\,R^{ab}_{\nu\mu}\,\delta
x^\mu\,\dot{x}^\nu\,,
\end{equation}
where in the last line we used an expression for the curvature
tensor that holds on $\Gamma$
$$
R^{ab}_{\nu\mu}(\sigma)=\bar{\omega}_{\nu,\mu\,\Gamma}-\bar{\omega}_{\mu,\nu\,\Gamma}.
$$
Equation (\ref{19}) is solved by
$$
\lambda^{ab}=-\delta
x^\mu\,\bar{\omega}^{ab}_{\mu\,\Gamma}+\tilde{\lambda}^{ab}=\tilde{\lambda}^{ab},
$$
where we used again the fact that $\bar{\omega}^{ab}_{\mu\,\Gamma}$
vanishes on $\Gamma$ and
$$
\tilde\lambda^{ab}(\sigma)=\int^\sigma
R^{ab}_{\nu\mu}(\sigma')\,\delta
x^\mu(\sigma')\,\dot{x}^\mu\,d\sigma'.
$$
Hence, the total variation reads
\begin{align}
\delta
X^a(\tau)&=\int_{\Gamma}d\sigma\,\tilde\lambda^a_{\,b}\,\bar{e}^b_\mu\,\dot
x^\mu\,+ \int_{\Gamma}d\sigma\,\left(\bar{e}_{\mu,\nu}\, \delta
x^\nu +\bar{e}_{\nu}^a\, \delta x^\nu{}_{,\mu}\right) \dot
x^\mu=\nonumber\\&=\int_{\Gamma}d\sigma\,\tilde\lambda^a_{\,b}\,\bar{e}^b_\mu\,\dot
x^\mu\,+\int_{\Gamma}d\sigma\,\left(\bar{e}_{\nu,\mu}\, \delta x^\nu
+\bar{e}_{\nu}^a\, \delta x^\nu{}_{,\mu}\right) \dot x^\mu
\nonumber\\&=\int_{\Gamma}d\sigma\,\tilde\lambda^a_{\,b}\,\bar{e}^b_\mu\,\dot
x^\mu\,+\int_{
\Gamma}d\sigma\,\frac{d}{d\sigma}\left(\bar{e}_{\nu}^a\, \delta
x^\nu\right)=\nonumber\\&=\bar{e}_{\nu}^a(x(\tau))\, \delta
x^\nu(x(\tau))+\int_{\Gamma}d\sigma\,\tilde\lambda^a_{\,b}\,\bar{e}^b_\mu\,\dot
x^\mu\,,
\end{align}
where we used $\delta x^\mu(t_1)=0$.

Notice that when $\delta
x^\mu(\tau)=x^\mu(\tau+d\tau)-x^\mu(\tau)=\dot x^\mu\,d\tau$,
$\tilde{\lambda}$ vanishes and one gets
\begin{equation}\label{7}
\frac{dX^a}{d\tau}=\bar{e}^a_\mu(x(\tau))\,\dot{x}^\mu\,.
\end{equation}
The total variation can be rewritten via an integration by part as
\begin{align}
\delta X^a=&\bar{e}_{\nu}^a(x(\tau))\, \delta
x^\nu(x(\tau))+\int_{\Gamma}d\sigma\,\tilde\lambda^a_{\,b}(\sigma)\,\dot{X}^b(\sigma)
\,=\nonumber\\
=&\bar{e}_{\nu}^a(x(\tau))\, \delta x^\nu(x(\tau))+\tilde\lambda^a_{\,b}(\tau)\,{X}^b(\tau)-\int_{\Gamma}d\sigma\,X^b(\sigma)\,\frac{d}{d\sigma}\tilde\lambda^a_{\,b}\,=\nonumber\\
=&\bar{e}_{\nu}^a(x(\tau))\,\delta x^\nu(x(\tau))+\int_0^\tau
d\sigma\,\left(X_b(\tau)-X_{b}(\sigma)\right)\,R^{ab}_{\mu\nu}
\,\delta x^\mu(\sigma)\,\dot{x}^\mu\,. \label{varfin}
\end{align}
This expression provides a linear map between the variations
$\delta x^\mu$ and $\delta X^a$. Since the basic physical variable
of the particle model is the position of the worldline $x^\mu(\tau)$,
and because the expression (\ref{varfin}) is very complex and
nonlocal, to make sure that the equations of motion following from
varying $X^a(\tau)$ and $x^\mu(\tau)$ are identical, we must show
that the linear mapping (\ref{varfin}) is invertible. This is done
in the Appendix A. Equations (\ref{7}) and (\ref{varfin}) contain
all the information we need to construct the action of a free
particle moving in curved spacetime and momentum space, and compute
the corresponding equations of motion from the variational
principle.

\subsection{The action and equations of motion}
Using the non-local variable $X^a$ discussed in the preceding
subsection we define the action of a particle moving in curved
spacetime and momentum spaces as follows
\begin{equation}\label{9}
   S=\int_{t_1}^{t_2} d\tau \left\{X^a[x(\tau)]\, E_a^\alpha\, \dot{p}_\alpha\,+N({\cal
   C}(p)\,-\,m^2)\right\}\,,
\end{equation}
where $X^a[x(\tau)]$ is calculated along the same curve $\Gamma$ as
the integral in (\ref{9}). Before turning to the discussion of the
properties of this action it is worth checking if it acquires the
desired form in the limiting cases of flat spacetime/momentum space,
respectively.

In the case of flat spacetime, to evaluate (\ref{5}) we have to find
the form of the associated Fermi tetrad. It follows from the tetrad
postulate (\ref{1f}) that since both $\omega_\mu{}^a{}_b$ and
$\Gamma^\alpha_{\mu\nu}$ vanish such tetrad must satisfy
$\partial_\mu\bar e^a_\nu=0$, and thus $\bar e^a_\mu=\delta^a_\mu$
(up to a global Lorentz transformation.) Then
$$
X^a = x^\mu(\tau)\,\delta^a_\mu - x^\mu(t_1)\,\delta^a_\mu\,.
$$
It can be checked that the second (constant) term produces neither a
contribution to the equation of motion nor a boundary term, and thus
for flat spacetime the action reproduces the one of Relative
Locality.

In the opposite case, when the momentum space is flat,
$E_a^\alpha=\delta_a^\alpha$. We integrate (\ref{9}) by parts and
use (\ref{7}) to obtain the standard curved spacetime particle
action (up to a boundary term), with the only difference being that now
we have to do with the Fermi tetrad instead of the generic one.
However, the equations of motion are the same in both cases, so we
may conclude that the actions are equivalent, the only difference
being that the Lagrangian (\ref{1}) is invariant under local Lorentz
symmetry, while in the action (\ref{9}) only the global Lorentz
symmetry remains\footnote{Notice also that, with the same caveat, the action~(\ref{9}) differs, up to a boundary term, from the action presented in~\cite{Kowalski-Glikman:2013xia} (written for Fermi tetrads) by the term $ \left( \int \dot{x}^\mu \bar{e}_\mu^a - x^\mu \bar{e}_\mu^a\right) \dot{E}^\alpha_a p_\alpha$.}.

Now we can turn to the equations of motion following from the action
(\ref{9}). Its variation reads
\begin{equation}
\delta S=\int_{t_1}^{t_2} d\tau \left(\delta X^{a}\,E_a^\alpha\,
\dot{p}_\alpha\,+X^{a}\,\delta(E_a^\alpha\,
\dot{p}_\alpha\,)+N\,\frac{\partial {\cal C}}{\partial
p_\alpha}\,\delta p_\alpha+\,\delta N\, ({\cal
C}(p)\,-\,m^2)\right)=0.
\end{equation}
Since we demonstrated that the map (\ref{varfin}) is invertible, we
know that the equations of  motion we get from the stationarity of
the action ($\delta S=0$) with respect to arbitrary variations
$\delta x^\mu$ are equivalent to the ones obtained by considering
arbitrary variations $\delta X^a$. Hence, we get
\begin{equation}\label{11}
E_a^\alpha\, \dot{p}_\alpha=0\,,\quad
\dot{X}^a\,E_a^\alpha=N\,\frac{\partial {\cal C}}{\partial
p_\alpha}\,,\quad {\cal C}(p)\,-\,m^2=0\,,
\end{equation}
which are equivalent to (for constant $N$)
\begin{equation}\label{12}
\dot{p}_\alpha=0,\qquad \ddot{X}^a=0.
\end{equation}
In particular, from the second relation in (\ref{12}) one finds
\begin{equation}\label{13}
\frac{d}{d\tau}{(\dot{x}^\mu\,\bar{e}^a_\mu)}=0\rightarrow
\ddot{x}^\mu+\Gamma^\mu_{\nu\rho}\,\dot{x}^\nu\,\dot{x}^\rho=0,
\end{equation}
where we used the expression of the Christoffel symbols in terms of
the tetrads $\bar{e}^a_\mu$, that follows from the tetrad postulate
(\ref{1f})
\begin{equation}\label{14}
\Gamma^\mu_{\nu\rho}=\bar{e}^\mu_a\,\partial_{(\rho}
\bar{e}^a_{\nu)}.
\end{equation}
Therefore, the trajectory in spacetime is a geodesic, independently
of the geometry in momentum space.

\section{Symmetries of the action}

Having discussed the form of the action, let us  now consider its
symmetries. First of all the action is manifestly invariant under
general coordinate transformations, in both momentum space\footnote{While the invariance under spacetime diffeomorphisms has a clear physical interpretation already in the flat momentum-space limit of action~(\ref{9}), the invariance under momentum space diffeomorphisms, which could be a mere formal invariance, has no classical counterpart, and we postpone its discussion to future studies. } and
spacetime, since $X^a$ is a spacetime scalar. 
Second it is invariant
under residual, global Lorentz transformations that leave invariant
the condition that the connection vanishes along the curve
$\omega|_\Gamma=0$.

From the Relative Locality perspective we are especially interested in translational symmetries: in the case of the model of a particle moving in
flat spacetime, the main features of relativity of spacetime locality are encoded in that the fact that the translations become momentum-dependent
 \cite{AmelinoCamelia:2011bm}, \cite{AmelinoCamelia:2011nt}, \cite{Kowalski-Glikman:2013rxa}.
As we will see an analogous effect takes place in the case of curved spacetime.

Like in the case of Relative Locality in flat spacetime
\cite{AmelinoCamelia:2011bm}, \cite{AmelinoCamelia:2011nt}, \cite{Kowalski-Glikman:2013rxa}, the
action (\ref{9}) is invariant (up to a boundary term) under the
translation
\begin{equation}\label{30}
    \delta X^a = E_\alpha^a(p)\, \xi^\alpha\,,\quad \dot
    \xi^\alpha=0\,.
\end{equation}
In flat-spacetime Relative Locality this symmetry translates rigidly
a (straight) particle worldline by an amount that depends on the
momentum carried by the particle. As we will see, an analogous
effect takes place in curved spacetime.

To see this we must find out what is the infinitesimal shift of the
particle trajectory $\delta x^\mu$ corresponding to the translation
(\ref{30}). Since we are interested in the effect that the
transformation (\ref{30}) has on trajectories, we assume that
the equations of motion are satisfied.

We start with (\ref{varfin})
\begin{equation}
E_\alpha^a(p)\, \xi^\alpha(\tau)=\bar{e}_{\nu}^a(x(\tau))\,\delta
x^\nu(x(\tau))+\int_{t_1}^\tau
d\sigma\,\left(X_b(\tau)-X_{b}(\sigma)\right)\,R^{ab}_{\mu\nu}
\,\delta x^\mu(\sigma)\,\dot{x}^\nu\,.\label{symm2}
\end{equation}
At $\tau=t_1$ we have
\begin{equation}\label{symm3}
E_\alpha^a(p)\, \xi^\alpha=\bar{e}_{\nu}^a(x(t_1))\,\delta
x^\nu(x(t_1))\,.
\end{equation}
This defines the first initial condition for $\delta x^\nu(x(t_1))$.
Next, let us differentiate (\ref{symm2}) over $\tau$
\begin{equation}\label{symm4}
    0=\frac{d}{d\tau}\left(\bar{e}_{\nu}^a(\tau)\,\delta x^\nu(\tau)\right)+\dot{X}_b(\tau)
\int_{t_1}^\tau d\sigma\,R^{ab}_{\mu\nu} \,\delta
x^\mu(\sigma)\,\dot{x}^\nu\,,
\end{equation}
so that at $\tau = t_1$
\begin{equation}\label{symm5}
    \bigg[\frac{d}{d\tau}\left(\bar{e}_{\nu}^a(\tau)\,\delta
    x^\nu(\tau)\right)\bigg]_{t_1}=0\,.
\end{equation}
Taking the second derivative of (\ref{symm2}) over $\tau$ we find
\begin{equation}\label{symm6}
 0   =\frac{d^2}{d\tau^2}\left(\bar{e}_{\nu}^a(\tau)\,\delta x^\nu(\tau)\right)+\ddot{X}_b(\tau)
\int_{t_1}^\tau d\sigma\,R^{ab}_{\mu\nu} \,\delta
x^\mu(\sigma)\,\dot{x}^\nu+\dot{X}_b(\tau) \,R^{ab}_{\mu\nu}(\tau)
\,\delta x^\mu(\tau)\,\dot{x}^\nu(\tau)\,.
\end{equation}
Since $\ddot X^a$ is zero by equations of motion, the second term in
the above expression disappears and we are left with
\begin{equation}
\bar e_{a\rho}\,\frac{d^2}{d\tau^2}\left(\bar{e}_{\nu}^a\,\delta
x^\nu\right)+R_{\mu\nu\rho\sigma} \,\delta x^\mu\,\dot{x}^\nu\,\dot
x^\sigma=0 ,\label{symm7}
\end{equation}
As shown in Appendix B this equation can be rewritten as
\begin{equation}\label{symm10}
\frac{D^2}{D\tau^2} \delta
x^\mu-R^\mu_{\,\,\nu\rho\sigma}\,\dot{x}^\nu\,\dot{x}^\rho\,\delta
x^\sigma=0\,,
\end{equation}
where $D/D\tau \equiv \dot x^\mu\nabla_\mu$ is the covariant
derivative projected along the worldline, subject to the initial
conditions
\begin{equation}\label{symm11}
   \delta x^\mu(t_1) =\bar{e}^{\mu}_a(x(t_1))\,E_\alpha^a(p)\,
   \xi^\alpha\,,\quad \frac{D}{D\tau} \delta x^\mu\bigg|_{t_1}=0\,.
\end{equation}
Equation (\ref{symm10}) is an equation of geodesic deviation and
therefore we see that the translational symmetry (\ref{30}) maps the
original geodesic, being the particle worldline, to another one, with
the magnitude of translation depending on the momentum carried by
the particle. This is exactly the effect one could foresee from the
flat spacetime Relative Locality, where straight lines (geodesics)
are translated by a constant, momentum dependent amount.

It follows from (\ref{symm11}) that $\delta x^\mu$ has the momentum
dependence encoded by the initial condition. Let us define another
variable $\zeta^\alpha$, which describe the momentum independent
translation
\begin{equation}\label{39}
\delta x^\mu =\bar{e}^{\mu}_a(x)\,E_\alpha^a(p)\,\zeta^\alpha\,.
\end{equation}
Since, as shown in Appendix B, both first and second covariant
derivatives of the tetrad $\bar e$ along the worldline  vanish, we
can rewrite (\ref{symm10}) as
\begin{equation}\label{40}
\frac{D^2}{D\tau^2} \zeta^\alpha-
\left(\bar{e}_{\mu}^a(x)\,E^\alpha_a(p)
R^\mu_{\,\,\nu\rho\sigma}\,\bar{e}^{\sigma}_b(x)\,E_\beta^b(p)
\right)\,\dot{x}^\nu\,\dot{x}^\rho\,\zeta^\beta=0\,,
\end{equation}
This equation describes a  congruence of particle worldlines in the
spacetime whose curvature is momentum-dependent. It might serve as a
starting point of more phenomenologically oriented investigations.
 
\section{Interactions}

\begin{figure}[h!]
\includegraphics[scale=0.6]{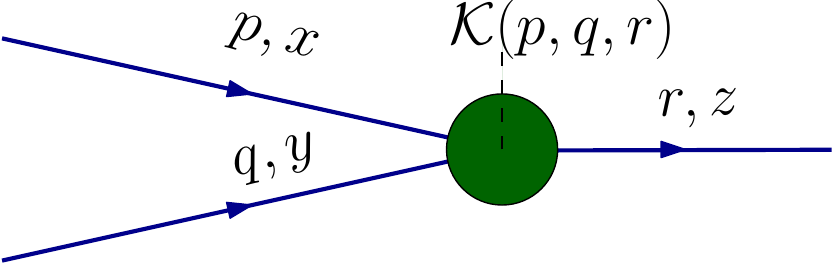}
\caption{A process with two incoming
particles of momenta $p$,$q$ and one outgoing particle of momentum $r$. ${\cal K}\left( q,p,r \right)$ is a function of the particles momenta enforcing energy-momentum conservation at the vertex. The graphic must not be intended as a spacetime representation but just as a qualitative picture illustrating the combination of momenta in the process.} \label{vertexFig}
\end{figure}

In the spirit of the Relative Locality framework introduced
in~\cite{AmelinoCamelia:2011bm} (see
also~\cite{AmelinoCamelia:2011nt} for an extensive discussion of the
properties of these boundary terms), in order to describe particle
processes (at a semi-classical, non quantum, level), we introduce in
the action~(\ref{3}) boundary terms enforcing constraints on the
endpoints of the particles worldlines. To illustrate how such a
constraint may be introduced in our framework, suppose that we want
to describe an idealized process depicted in Fig.~\ref{vertexFig},
with two incoming particles labeled respectively by momenta and
coordinates $p_\alpha,x^\mu$, $q_\alpha,y^\mu$ and the outgoing one
labeled by $r_\alpha, z^\mu$.

\noindent
The action for this process is
\begin{align}
S =& \int_{-\infty}^{t}d\tau\left[X^{a}E_a^\alpha\dot{p}_{\alpha} +
N_p\left({\cal C}(p) - m_p^{2}\right) \right]
\nonumber\\
+&\int_{-\infty}^{t}d\tau\left[ Y^{a}E_a^\alpha\dot{q}_{\alpha} +
N_q\left({\cal C}(q) -
m_q^{2}\right)\right]\nonumber\\
+&\int^{\infty}_{t}d\tau\left[Z^{a}E_a^\alpha\dot{r}_{\alpha} +
N_r\left({\cal C}(r) - m_r^{2}\right) \right]- k^{\alpha}{\cal
K}_{\alpha}\left(p,q,r\right) \bigg|_{\bar{\tau}}\,.
\label{actionInt}
\end{align}
In this formula $k^\alpha$ is a Lagrange multiplier enforcing the
constraint  ${\cal K}_{\alpha}\left(p,q,r\right)$ on the worldlines
endpoints, which plays a role of a (in general) deformed law of
energy-momentum conservation at the vertex. Typically, ${\cal
K}_{\alpha}\left(p,q,r\right) = \left( p \oplus q \oplus
(\ominus r)\right)_\alpha$, with the symbol $\oplus$ ($\ominus$) encoding the connection
on the momentum space geometry, characterizing the (in general
non-linear) law of summation for
momenta~\cite{AmelinoCamelia:2011bm}.  In addition to the equations
of motion for the bulk part Eqs.~(\ref{11}), (\ref{13}), the
boundary term contributes with the constraints
\begin{gather}
{\cal K}_{\alpha}\left(p,q,r\right)\big|_{t}= \left( p \oplus q \oplus
(\ominus r)\right)_\alpha \big|_{t} =0, \label{constraint1} \\
X^{a}\left( t \right) =  k^{\beta} E^a_\alpha (p) \frac{\partial{\cal K}_{\beta}}{\partial p_{\alpha}}\big|_{t}, \label{constraint2}\\
Y^{a}\left( t \right) =  k^{\beta} E^a_\alpha (q)
\frac{\partial{\cal K}_{\beta}}{\partial q_{\alpha}}\big|_{t}.
\label{constraint3}\\
Z^{a}\left( t \right) =  k^{\beta} E^a_\alpha (r)
\frac{\partial{\cal K}_{\beta}}{\partial r_{\alpha}}\big|_{t}.
\label{constraint4}
\end{gather}

When $k^\alpha$ changes, the $X^a$ transforms as
\begin{equation}
\delta X^{a}\big|_{t}=E_{\alpha}^{a}(p)\frac{\partial\left( p \oplus q \oplus
(\ominus r)\right)_\beta}{\partial p_{\alpha}}\delta k^{\beta}\big|_{t},
\label{varConstraint}
\end{equation}
with analogous relations holding for the other particles.
Assuming that we take the initial condition for the geodesic
deviation equation (\ref{symm11}) at the interaction point, we find
that
\begin{equation}
\delta x^{\mu}\big|_{t}=\bar e^\mu_a\,
E_{\alpha}^{a}(p)\frac{\partial\left( p \oplus q \oplus
(\ominus r)\right)_\beta}{\partial p_{\alpha}}\delta k^{\beta}\big|_{t}\,.
\end{equation}
We see therefore that the structure of the interaction vertex in the
case of curved spacetime is essentially the same as in the flat
spacetime case of Relative Locality~\cite{AmelinoCamelia:2011bm}.

\section{Conclusions}

In this paper we generalized Relative Locality, originally
defined~\cite{AmelinoCamelia:2011bm} in flat spacetime, to the case
of an arbitrary, curved background spacetime, preserving invariance under general coordinate transformations. It turns out that on
the formal level this latter theory is a natural generalization of
the former one: free particle trajectories are now geodesics
instead of flat-spacetime straight lines, and rigid, momentum dependent
translations of the flat case are replaced with geodesic deviations,
sensitive to curvature.

In spite of the apparent similarities there are, however, some major
differences between flat and curved spacetime case. In the
latter we were forced to use nonlocal variables $X^a$ to define the
action that had desired symmetry properties. This might be just a
technical artifact, but it may also signal  a presence of some
deeper layer present in theories with a nontrivial geometry in both
momentum space and spacetime. Furthermore, in flat spacetime the 
symmetries of the action are associated with some transformations (rigid translations) defined in the whole spacetime manifold. On the contrary, 
in the curved case we have to solve the equation of geodesic deviation 
on a given geodesic to find the symmetry. This implies that the transformation 
$\delta x^\mu$ which leaves the action invariant (up to a boundary term) depends on each 
particular solution of equations of motion and generically, because of curvature, it cannot 
be extended to the whole spacetime.  

Aside from its conceptual relevance, our result opens some interesting phenomenological perspectives.
Indeed most of the opportunities to test Planck-scale deformation effects on particle kinematics, that have been proposed in the recent literature, rely on some source of amplification of the relevant effects due to cosmological distance of astrophysical sources. 
Most of the results\footnote{In~\cite{AmelinoCamelia:2012it}, a first investigation of the interplay between spacetime expansion and relativity of locality has been presented, for the case of de Sitter-like spacetime expansion.} for theories with curved momentum space (and earlier of the DSR
theories) have been discussed in the context of flat spacetime, while in the proposed scenarios relevant for Planck-scale phenomenology the effect of spacetime curvature cannot be neglected.
Our result can be taken then as starting point for further studies of Relative Locality effects in presence of spacetime curvature. Moreover using the results of the
present work one may try to investigate the Relative Locality
effects in the case of strong gravitational field, for example in
the context of black hole physics (see \cite{Smolin:2011ns}).

\section*{Acknowledgment}

We would like to thank Laurent Freidel and Lee Smolin for
discussion. For FC, JKG, and GR this work was supported by funds
provided by the National Science Center under the agreement DEC-
2011/02/A/ST2/00294. For JKG this work was also supported in parts
by the grant 2011/01/B/ST2/03354.

\appendix

\section{Invertibility of $\delta X^a$}

To show that the map (\ref{varfin})  is invertible  we must show
that its kernel contains only $\delta x^\mu=0$, to wit
\begin{equation}
\delta X^a=0\Rightarrow\delta x^\mu=0\,, \label{0kernel}
\end{equation}
 In order to prove this let us note that
since we assumed $x^\mu=x^\mu(\tau)$ to be a $C^\infty$ function of
$\tau$, $X^a(\tau)$ and $\delta X^a(\tau)$ are $C^\infty$ too and
the condition $\delta X^a(\tau)=0$ for each $\tau$ is equivalent to
\begin{equation}
\bigg[\frac{d^n}{d\tau^n}\delta X^a(\tau)\bigg]\bigg|_0=0,\qquad
\forall n\in\mathbb{N}\,.
\end{equation}
From the expression (\ref{varfin}) we have
\begin{equation}
\frac{d}{d\tau}\delta
X^a(\tau)=\frac{d}{d\tau}\left(\bar{e}_{\nu}^a(x(\tau))\,\delta
x^\nu(x(\tau))\right) +\frac{dX_b}{d\tau}\int_0^\tau
d\sigma\,R^{ab}_{\mu\nu} \,\delta x^\mu(\sigma)\,\dot{x}^\mu,
\end{equation}
and for $\tau=0$ we get
\begin{equation}
0=\bigg[\frac{d}{d\tau}\delta
X^a(\tau)\bigg]\bigg|_0=\bigg[\frac{d}{d\tau}\left(\bar{e}_{\nu}^a\,\delta
x^\nu\right)\bigg]\bigg|_0\,,
\end{equation}
which implies (we remind that $\delta x^\mu(0)=0$)
\begin{equation}
\bigg[\frac{d}{d\tau}\delta x^\mu\bigg]\bigg|_0=0\,.
\end{equation}
Similarly, one can show that
\begin{equation}
\bigg[\frac{d^2}{d\tau^2}\delta
X^a(\tau)\bigg]\bigg|_0=0\Rightarrow\bigg[\frac{d^2}{d\tau^2}\delta
x^\mu\bigg]\bigg|_0=0\,,
\end{equation}
and by iterating one gets
\begin{equation}
\bigg[\frac{d^n}{d\tau^n}\delta x^\mu(\tau)\bigg]\bigg|_0=0,\qquad
\forall n\in\mathbb{N}\,,
\end{equation}
which is equivalent to $\delta x^\mu(\tau)=0$ for each
$\tau\in[t_1,t_2]$. Therefore we proved (\ref{0kernel}) and the map
(\ref{varfin}) is invertible.

\section{Derivation of geodesic deviation}

In this appendix we show that eq.\ (\ref{symm7}) is equivalent to
the equation of geodesic deviation (\ref{symm10}). To see this
notice that
\begin{equation}
\delta \dot{x}^a=\frac{d}{d\tau}(\delta
x^\mu\,\bar{e}^a_\mu)=\left(\frac{D}{D\tau}\delta
x^\mu\right)\,\bar{e}^a_\mu+ \delta
x^\mu\,\frac{D}{D\tau}\bar{e}^a_\mu\,,
\end{equation}
and
\begin{equation}
\delta \ddot{x}^a=\left(\frac{D^2}{D\tau^2}\delta
x^\mu\right)\,\bar{e}^a_\mu+ \delta
x^\mu\,\frac{D^2}{D\tau^2}\bar{e}^a_\mu+2\left(\frac{D}{D\tau}\delta
x^\mu\right)\,\frac{D}{D\tau}\bar{e}^a_\mu\,.
\end{equation}
It follows from the tetrad postulate (\ref{1f}) and the properties
of Fermi tetrads that
\begin{equation}\label{b1}
\frac{D}{D\tau}\bar{e}^a_\mu=0\,.
\end{equation}
As for the second covariant derivative we get
$$
\frac{D^2}{D\tau^2}\bar{e}^a_\mu=\frac{D}{D\tau}\left(\dot{x}^\nu\,\bar{e}^b_{\mu}\right)\,\omega^{a}_{\,\,b\nu}+
\dot{x}^\nu\,\bar{e}^b_{\mu}\,\frac{D}{D\tau}\omega^{a}_{\,\,b\nu}\,.
$$
This expression is again zero for a Fermi tetrad, because connection
$\omega$ is zero everywhere on the worldline and thus its derivative
along it vanishes; therefore
\begin{equation}\label{b2}
\frac{D^2}{D\tau^2}\bar{e}^a_\mu=0\,.
\end{equation}
Using (\ref{b1}) and (\ref{b2}) one straightforwardly derives
(\ref{symm10}).

\end{document}